\documentclass[sigconf,nonacm]{acmart}

\AtBeginDocument{%
  }

\copyrightyear{2025}
\acmYear{2025}
\setcopyright{cc}
\setcctype{by}
\acmConference[ISSTA Companion '25]{34th ACM SIGSOFT International Symposium on Software Testing and Analysis}{June 25--28, 2025}{Trondheim, Norway}
\acmBooktitle{34th ACM SIGSOFT International Symposium on Software Testing and Analysis (ISSTA Companion '25), June 25--28, 2025, Trondheim, Norway}\acmDOI{10.1145/3713081.3731735}
\acmISBN{979-8-4007-1474-0/2025/06}

\def\name{{\sc InfraFix}}
\newcommand{\TODO}[1]{\textcolor{red}{#1}\GenericWarning{}{LaTeX Warning: TODO: #1}}\newcommand\todo\TODO

\usepackage{makecell}
\usepackage[inline]{enumitem}

\begin{document}

\title{\name: Technology-Agnostic Repair of Infrastructure as Code}

\author{Nuno Saavedra}
\email{nuno.saavedra@tecnico.ulisboa.pt}
\affiliation{%
  \institution{INESC-ID/IST, University of Lisbon}
  \city{Lisbon}
  \country{Portugal}
}

\author{João F. Ferreira}
\email{joao@joaoff.com}
\affiliation{%
  \institution{INESC-ID/IST, University of Lisbon}
  \city{Lisbon}
  \country{Portugal}
}

\author{Alexandra Mendes}
\email{alexandra@archimendes.com}
\affiliation{%
  \institution{INESC TEC, Faculty of Engineering, University of Porto}
  \city{Porto}
  \country{Portugal}
}

\renewcommand{\shortauthors}{Saavedra et al.}

\begin{abstract}
Infrastructure as Code (IaC) enables scalable and automated IT infrastructure management but is prone to errors that can lead to security vulnerabilities, outages, and data loss. While prior research has focused on detecting IaC issues, Automated Program Repair (APR) remains underexplored, largely due to the lack of suitable specifications. In this work, we propose \name, the first technology-agnostic framework for repairing IaC scripts. 
Unlike prior approaches, \name\ allows APR techniques to be guided by diverse information sources.
Additionally, we introduce a novel approach for generating repair scenarios, enabling large-scale evaluation of APR techniques for IaC. We implement and evaluate \name\ using an SMT-based repair module and a state inference module that uses system calls, demonstrating its effectiveness across 254,288 repair scenarios with a success rate of 95.7\%. Our work provides a foundation for advancing APR in IaC by enabling researchers to experiment with new state inference and repair techniques using \name\ and to evaluate their approaches at scale with our repair scenario generation method.
\end{abstract}




\maketitle

\section{Introduction}

Infrastructure as Code (IaC) is the process of managing IT infrastructure via programmable configuration files. As IT infrastructure scales, the cost of maintenance and the risk of manual misconfigurations due to human error drastically increases. 
It becomes impracticable to fully manage IT infrastructures manually and, for that reason, IaC has progressively gained more adoption. 

Despite the benefits provided by IaC, such as faster, more reliable, and predictable deployments, errors in these critical pieces of code can lead to security vulnerabilities~\cite{rahman2019seven,rahman2021security,reis2022leveraging,saavedra2022glitch,saavedra2023polyglot,opdebeeck2023control}, application outages~\cite{gitlab_outage_2014,stack_exchange_outage_2014,reddit_outage_2016,incident_resolution_honeycomb_2021,rust_outage_2023}, and data losses~\cite{wikimedia_dataloss_2016}, which undermine the stability and reliability of 
services and ultimately lead to financial losses or even the loss of human lives. For example, in October 2021, aid workers in conflict zones like Syria were using WhatsApp to communicate about where bombings were taking place, so they could move safely through the country~\cite{facebook_down_2021}. However, during Facebook’s outage, which was triggered by a misconfiguration in their internal backbone network that disconnected Facebook's services and tools~\cite{Janardhan_2021}, the aid workers were unable to communicate.

Given this, it is crucial to understand how we can detect and prevent errors in IaC. Currently, the IaC research community is primarily focused on error detection~\cite{saavedra2022glitch,saavedra2023polyglot,rahman2021security,reis2022leveraging,hassan2024state,begoug2024fine,sotiropoulos2020practical,opdebeeck2022smelly,rahman2023security,lepiller2021analyzing,schwarz2018code,sharma2016does,bessghaier2024prevalence,opdebeeck2023control,zerouali2023helm}. Only a few studies have taken the additional step of exploring Automated Program Repair (APR) in IaC~\cite{weiss2017tortoise,hassan2018rudsea}, and most of these efforts focus on static repair rules that do not depend on program-specific specifications~\cite{bui2023dockercleaner,henkel2021shipwright,durieux2023parfum}. Since most APR techniques use weak specifications, such as test suites~\cite{goues2019automated,repair-living-review, monperrus2018automatic}, 
the development of APR in IaC is being hindered by the lack of adoption of testing practices for IaC, 
as noted in prior work
\cite{guerriero2019adoption,sokolowski2024automated,rahman2019systematic,sokolowski2024pipr}.

In light of IaC's limited testability, two potential avenues to enable advancements in APR for this domain are:
\begin{enumerate*}
\item exploring and improving testing practices for IaC\,---\,some research has been conducted in this direction~\cite{hasan2020testing,sokolowski2024automated,sokolowski2024pipr};
\item finding different types of specifications for IaC scripts that APR techniques can rely on. 
\end{enumerate*}
We focus our attention on the second avenue. Previous works have presented different types of specifications for IaC~\cite{hassan2018rudsea,weiss2017tortoise}. In one of these works, Weiss et al. developed a technique called \emph{imperative configuration repair}, which enables automatic repair of IaC scripts guided by a series of shell commands~\cite{weiss2017tortoise}. Since a common workflow adopted by sysadmins involves using the shell to identify and test required changes to IaC scripts, Weiss et al. used these commands as the specification that guides their APR technique. They implemented their technique in a tool called Tortoise~\cite{weiss2017tortoise}.

Building on these ideas, we propose \name, a technology-agnostic framework for repairing IaC scripts. While previous approaches rely on specific information sources, such as shell commands, as specifications, \name\ is more general by allowing researchers to develop APR techniques guided by specifications derived from diverse information sources. Both the APR technique used and the sources can be adapted by modifying the repair and state inference modules, respectively. 
Moreover, \name\ overcomes the limitations of previous APR approaches that were not designed to support multiple IaC technologies simultaneously. According to the 2024 Stack Overflow Developer Survey~\cite{Overflow_2024}, the IaC landscape is highly fragmented, with adoption rates that vary significantly between technologies. For example, even though Ansible (7.9\%) and Terraform (10.6\%) are widely used technologies, others such as Puppet (1.2\%), Chef (0.7\%) and Nix (2.7\%) still have considerable adoption rates. Given this diversity, a technology-agnostic approach is highly desirable.

We make the following contributions:
\begin{enumerate*}
    \item We propose \name, the first technology-agnostic framework that can automatically repair IaC scripts. Our implementation includes an example of an SMT-based repair module and a state inference module.
    \item We introduce a novel approach for the generation of repair scenarios that can be used to evaluate APR approaches for IaC. We use this approach to evaluate our implementation of \name\ in 254,288 different repair scenarios, achieving a successful repair rate of 95.7\%.
\end{enumerate*}

\textbf{Tool Availability.} A replication package with all the datasets of IaC scripts used to create repair scenarios, the scripts used to run our generation approach, and the source code of the implementation of \name\ is available at: \url{https://doi.org/10.5281/zenodo.15061764}. A video describing and demonstrating \name\ is available at: \url{https://youtu.be/ETSW0n9DcdU}. \name\ is open-source and available at: \url{https://github.com/sr-lab/GLITCH/tree/issta2025}.

\section{InfraFix}

\begin{figure}
    \centering
    \includegraphics[width=\linewidth]{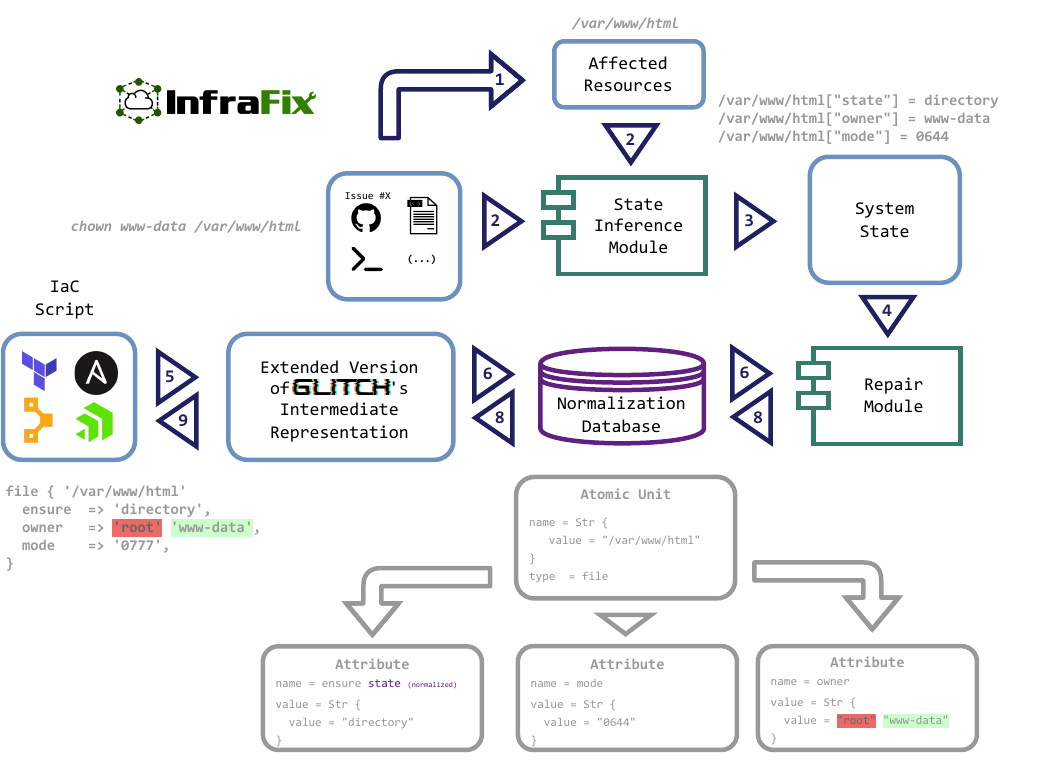}
    \caption{Overview of \name's architecture.}
    \label{fig:overview}
\end{figure}

This section presents \name's technology-agnostic design and modules. Figure~\ref{fig:overview} shows an overview of \name's architecture. \name\ starts by taking as input information that describes the desired state of the configured system and then infers the corresponding system state from it (\S~\ref{sec:state_inference}). The framework then receives as input the IaC script to be repaired, which can currently be written in Ansible, Puppet, Chef, or Terraform, and translates it into a normalized intermediate representation (IR) (\S~\ref{sec:tech-agnostic}). Finally, the normalized IR is repaired according to the desired system state, and the changes are propagated back to the original IaC script (\S~\ref{sec:repair}).

\subsection{State Inference Module}
\label{sec:state_inference}

A state inference module creates the specification that guides the repair of the IaC script. This module receives as input information about the system, such as logs or system calls, and a list of resources affected by that information and infers the desired system state.
A \emph{system state} is a list of resource states. A resource state consists of the resource's identifier and a dictionary with the attributes of the resource. 
Due to the declarative nature of IaC scripts, system states are a suitable specification to guide APR techniques.
We can use different sources of information to infer states, such as system calls from executed shell commands~\cite{weiss2017tortoise}, cloud logs such as AWS CloudTrail or Google Cloud Audit Logs, or GitHub issues. 

The current implementation of \name\ extends Weiss et al.'s approach, providing a state inference module that uses system calls to get a list of files affected by a sysadmin's shell commands. The module then uses that list to retrieve the attributes of those files from the file system.
Importantly, \name\ is flexible and can be easily extended with other valuable approaches.
For example, consider the scenario where a sysadmin is asked to install Steam\footnote{Steam is a video game digital distribution service managed by Valve.} on all Linux machines in a gaming convention. The sysadmin develops an Ansible script that installs the \emph{steam} package on these machines. However, when trying to deploy the script, it returns the following error:

\begin{scriptsize}
\begin{verbatim}
Error: You are missing the following 32-bit libraries, and Steam may not run:
libGL.so.1  libdrm.so.2
\end{verbatim}
\end{scriptsize}

This error is a real-world error that affected many Linux users~\cite{steam_issue_2025}. Given this error message, a state inference module can produce the following system state:
\begin{center}
\begin{scriptsize}
\begin{tabular}{cc}
package:steam \{ state $\rightarrow$ present & package:libgl1-mesa-dri:i386 \{ state $\rightarrow$ present \\ \multicolumn{2}{c}{package:libgl1:i386 \{ state $\rightarrow$ present}\\
\end{tabular}
\end{scriptsize}   
\end{center}

The inferred system state ensures that the packages required to overcome the error are installed. This system state could guide an APR technique to fix the original Ansible script.


\subsection{Technology-Agnostic Approach}
\label{sec:tech-agnostic}

To ensure \name\ is technology-agnostic, we need to address two main challenges:
\begin{enumerate*}
    \item \textbf{Syntactic structure:} Different IaC technologies may use different configuration languages. Dealing simultaneously with different configuration languages requires an abstraction that overcomes the different syntactic structures.
    \item \textbf{Naming:} Depending on the IaC technology, attributes, reserved values, or resource types with the same semantic meaning may have different names. For instance, in Chef, links and directories are managed using distinct resources, whereas in Ansible and Puppet, files, links, and directories are all handled within the same \emph{file} module.
\end{enumerate*}

Overcoming these two aspects enables APR techniques to operate consistently across technologies. For example, once the naming is unified, APR can apply the same transformations to the \emph{file} module in Ansible and Puppet as in the \emph{link} and \emph{directory} modules in Chef. Similarly, abstracting control structures such as conditionals and blocks allows to treat them uniformly.

\subsubsection{GLITCH's Intermediate Representation} 

\name\ addresses the challenge of syntactic differences by leveraging GLITCH's Intermediate Representation (IR).
GLITCH is a framework that enables automated technology-agnostic code smell detection for IaC scripts~\cite{saavedra2022glitch,saavedra2023polyglot}.
GLITCH employs an IR into which the scripts of multiple technologies are transformed, and on which analysis techniques are defined to identify code smells.
GLITCH's IR follows an object-oriented approach with a hierarchical structure and is able to capture similar concepts from different IaC technologies. 

To support APR, we extended GLITCH's IR to support structured expressions. GLITCH's original IR encoded all expressions as a single string. For example, a sum previously represented as the string \emph{"x + y"} is now expressed in our extended IR as a \emph{Sum} node containing two \emph{Variable Reference} nodes. This allows APR techniques to manipulate the values of expressions and pinpoint the location of values in the original IaC script to modify their textual representation. Figure~\ref{fig:overview} shows a graph-based visualization of how our GLITCH's extended IR models a snippet of an IaC script. For a more detailed description of GLITCH's IR, we refer to GLITCH's original paper by Saavedra and Ferreira~\cite{saavedra2022glitch}.

\subsubsection{Normalization Database}

\name\ uses a database to normalize resource types, attribute names, and reserved attribute values based on their semantics. 
If two types of resources in different technologies exhibit the same behavior, they will be mapped to the same type, meaning they will share the same string in the normalized intermediate representation.
The same happens between two semantically equivalent pairs of attribute names and values. For example, in Figure~\ref{fig:overview}, the attribute \emph{ensure} is normalized to \emph{state}, its equivalent in Ansible. The chosen normalization names are arbitrary, but must remain consistent across technologies.

The normalization database reduces the cost of supporting semantically equivalent functionality across different technologies. For deterministic APR techniques, it eliminates code duplication. For machine learning APR techniques, training on the normalized IR enables models to possibly generalize across supported technologies. It also expands the available data, allowing scripts from multiple technologies to be used for both training and evaluation.

While only standard system resources may seem to justify normalization, common configuration needs across IaC tools create natural overlaps, making the normalization process broadly applicable.
For example, the \emph{community.aws.lightsail} module in Ansible~\cite{AnsibleLighsail_2025} overlaps significantly with the \emph{aws\_lightsail\_instance} resource in Terraform~\cite{TerraformLighsail_2025}.

\subsection{Repair Module}
\label{sec:repair}

The repair module takes as input the normalized IR of an IaC script and a system state, and is responsible for repairing the normalized IR to align with the desired specification.
The repair is applied to the normalized IR and then propagated back to the original IaC script. This process first reverses the normalization rules and then uses information about the location of modified components in the IR, namely the line and column positions, to update their textual representation in the original script.

We integrate the SMT-based approach of Weiss et al.~\cite{weiss2017tortoise} into~\name, but we extend it to support repairs for other types of resources other than files. We adopt this approach as it represents the state-of-the-art in APR for IaC.
Currently, this module supports the repair of the following resources: files, packages, services and users in Ansible, Puppet, and Chef; AWS IAM roles, AWS EC2 instances, and AWS S3 buckets in Terraform and Ansible. We selected these types of resources because they are shared across the supported IaC technologies and rank among the most frequently used, as evidenced by the statistics provided in our replication package.
Furthermore, we improved the approach of Weiss et al. by incorporating support for conditional branching, attribute sketching~\cite{solar2006combinatorial}, and direct handling of attributes without requiring preprocessing of IaC scripts, thus expanding the spectrum of potential repairs. 

\section{Evaluation}
In this section, we describe the datasets used to evaluate \name, introduce our automated repair scenario generation method, and assess \name’s effectiveness in repairing IaC scripts.

\subsection{Datasets}

\begin{table}[]
    \centering
    \scriptsize
    \caption{Dataset metrics.}
    \label{tab:datasets}
    \begin{tabular}{rccccc}
        \toprule
        & \textbf{Ansible} & \textbf{Puppet} & \textbf{Chef} & \textbf{Terraform} & \textbf{Tortoise} \\
        \midrule
        All scripts                                   & 108,511 & 17,311 & 70,942 & 41,550 & 13 \\
        Without test and library scripts              & 99,934  & 15,725 & 40,372 & 40,972 & 13 \\
        Without duplicates                            & 69,524  & 12,973 & 19,846 & 31,687 & 13 \\
        \makecell[r]{\textbf{Curated dataset} \\ (Parsable by \name)}  
                                                     & 64,011  & 12,753 & 19,716 & 30,566 & 13 \\
        \textbf{Benchmark}                            & 12,358  & 4,868  & 3,568  & 1,300  & 13 \\
        \bottomrule
    \end{tabular}
\end{table}

To evaluate \name, we use
the Ansible and Chef datasets provided by Saavedra and Ferreira~\cite{saavedra2022glitch}, the Puppet datasets provided by Rahman et al.~\cite{rahman2019seven}, and the Terraform dataset provided by Gonçalves~\cite{goncalves2023}. We use these datasets since they have been previously used for large-scale empirical studies on IaC.
We also included in our evaluation the Puppet files used to evaluate Tortoise~\cite{weiss2017tortoise}.
Subsequently, we curated the IaC scripts in the selected datasets based on the following criteria:
\begin{enumerate*}
    \item Test or library scripts are discarded. 
    \item If two files contain the same content, only one is kept.
    \item Files that are not parsable by \name\ due to errors are discarded.
\end{enumerate*}
After applying all the criteria, we obtain five curated datasets. To obtain our evaluation benchmarks, we filter out all files that do not contain at least one resource currently supported by \name.
Table~\ref{tab:datasets} shows the number of scripts in the datasets after each step. 

\subsection{Automated Generation of Repair Scenarios}

\begin{figure}
    \centering
    \includegraphics[width=0.7\linewidth]{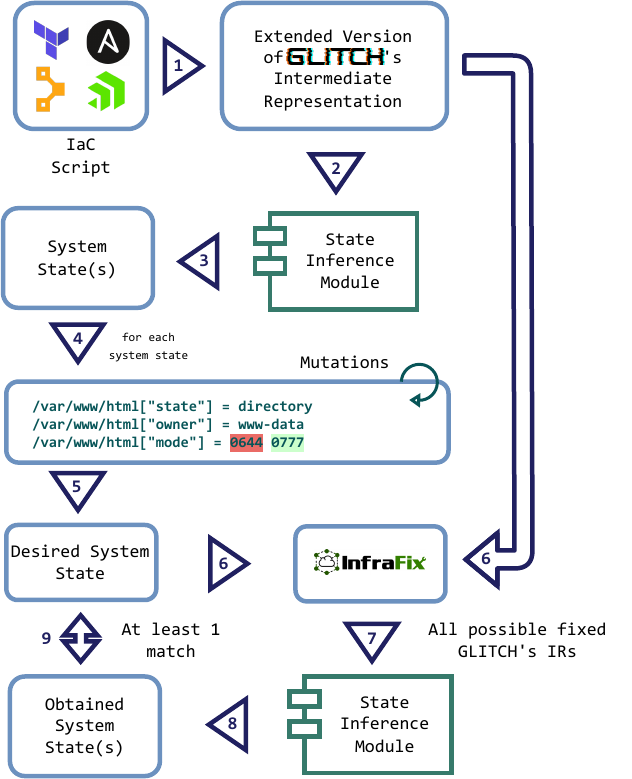}
    \caption{Overview of the method for automated evaluation of APR approaches for IaC.}
    \label{fig:evaluation}
\end{figure}

To perform a comprehensive evaluation of \name, we need to generate repair scenarios for each file in our benchmarks.
Manually generating repair scenarios is time-consuming, rendering it unfeasible for large-scale empirical evaluations. For this reason, we created a novel method to automatically generate them. Figure~\ref{fig:evaluation} shows an overview of our method.

First, we transform the IaC script into the normalized GLITCH's IR.
Then, we employ a state inference module that takes the IR as input to determine the possible system state(s) resulting from applying the original IaC script. We note that multiple potential system states may arise when it is not possible to statically determine which branch of a conditional statement will be executed.
After obtaining the possible system states, we apply a sequence of mutations to each state, obtaining a set of desired system states.
These mutations simulate potentially desired modifications to the system state.
Mutations involve modifying the attributes of the resources in the state of the system with arbitrary yet valid values. 

Each desired system state corresponds to a repair scenario.
For each scenario, we execute \name\ with the normalized IR and the desired system state. Since \name\ can generate multiple possible solutions, we consider the repair successful if any of the fixed GLITCH's IR results in a system state equivalent to the desired one. If \name\ raises an error, the scenario is marked as erroneous. If there is a timeout without finding any solutions, we consider the repair scenario to be timed out. Otherwise, we consider it failed. The timeout threshold is set to 120 seconds.

\subsection{Results}

\begin{table}
    \centering
    \caption{Repair scenarios' results for each benchmark.}
    \label{tab:test_results}
    \scriptsize
    \begin{tabular}{rccccc}
        \toprule
                 & \textbf{Total} & \textbf{Passed (\%)} & \textbf{Failed (\%)} & \textbf{Error (\%)} & \textbf{Timeout (\%)}  \\
        \toprule
         \textbf{Ansible}  & 183,825 & 176,121 (95.8 \%)  & 256 (0.1 \%)    & 155 (0.1 \%) &  7,293 (4.0 \%) \\
         \textbf{Puppet}   & 4,401   & 3,497 (79.5 \%)    & 852 (19.4 \%)   & 0 (0.0 \%)   & 52 (1.2 \%) \\
         \textbf{Chef}     & 61,357  & 59,181 (96.5 \%)   & 1,130 (1.8 \%)  & 64 (0.1\%)   & 982 (1.6 \%) \\
         \textbf{Terraform}& 3,544   & 3,544 (100.0 \%) & 0 (0.0\%) & 0 (0.0\%) & 0 (0.0\%) \\
         \textbf{Tortoise} & 1,161   & 1,064 (91.7\%)     & 80 (6.9 \%)    & 0 (0.0\%)    & 17 (1.5\%) \\
        \midrule
         \textbf{Total}    & 254,288 & 243,407 (95.7\%)   & 2,318 (0.9\%)   & 219 (0.1\%) & 8,344 (3.3 \%) \\
        \bottomrule
    \end{tabular}
\end{table}

We successfully generated repair scenarios for 7,833/12,358 (Ansible), 125/4,868 (Puppet), 2,069/3,568 (Chef), 736/1,300 (Terraform), and 13/13 (Tortoise) files.
Failures were due to two main reasons: (1) although these files contained supported resources, their names could not be statically resolved to strings; and (2) the resources were defined within other constructs, such as function calls or classes, that were not invoked or instantiated within the same script. As a result, these resources could not be included in the inferred system state for those files.

We generated 183,825 repair scenarios for Ansible, 4,868 for Puppet, 61,357 for Chef, 3,544 for Terraform, and 1,161 for Tortoise. The experiments were conducted on a machine with an Intel Xeon Silver 4110 CPU @ 2.10GHz and 64GB of RAM. We used 32 cores to parallelize the evaluation with all repair scenarios of a single file being handled by the same process.
Table~\ref{tab:test_results} shows the results obtained for each benchmark. 
\name\ passed 95.8\%, 79.5\%, 96.5\%, 100.0\%, and 91.4\% of the repair scenarios in the Ansible, Puppet, Chef, Terraform, and Tortoise benchmarks, respectively.
Upon manually examining the repair scenarios, we found most failures were due to: (1) references to undefined variables assigned to attributes or other variables, (2) attributes containing shared variable references that, when modified, unintentionally affect multiple attributes, and (3) defining multiple resources within the same construct using an array as title~\cite{array_titles_puppet_2025}.

We do not compare with Tortoise, as it is deprecated due to missing dependencies. We contacted the first and second authors of Tortoise~\cite{weiss2017tortoise}, who confirmed that these dependencies are no longer available, making it impossible to compare Tortoise with new APR tools and techniques for IaC. Future work in this area requires a new state-of-the-art tool, for which we propose \name.

\section{Related Work}


Due to space constraints, we focus on repair techniques specific to IaC.
Bui et al.~\cite{bui2023dockercleaner} and Durieux et al.~\cite{durieux2023parfum} developed tools for automated detection and repair of code smells in Dockerfiles. 
These works are similar to ours in their attempt to repair IaC scripts. 
However, they focus on code smells, while we consider a wider range of issues.
Henkel et al. used machine learning to infer repair rules based on build log analysis. They developed a tool, called Shipwright, with the inferred rules to repair Dockerfiles that fail to build~\cite{henkel2021shipwright}. 
The main difference is that Shipwright uses pre-defined and inferred rules, while \name\ infers system state to guide repairs.
Hassan et al. proposed RUDSEA, an approach that automatically recommends updates to Dockerfiles based on changes in the source code and build configuration files~\cite{hassan2018rudsea}. Weiss et al. proposed Tortoise, a tool that implements a technique called \emph{imperative configuration repair}, which enables automatic repair of IaC scripts guided by a series of shell commands~\cite{weiss2017tortoise}.
Similarly to RUDSEA, Tortoise is focused on a single IaC technology. Our approach is more general, both in terms of resources and IaC technologies supported. 
In fact, the specifications used by RUDSEA
could be used with 
\name,
provided that a state inference module is implemented.

\section{Conclusion}
We introduce \name, the first technology-agnostic framework for APR in IaC. 
Beyond supporting multiple IaC technologies, \name\ enables researchers to explore diverse state inference and repair techniques. 
We also propose a novel
method for generating repair scenarios that facilitate large-scale evaluation of APR solutions for IaC. To demonstrate \name’s capabilities, we implement an SMT-based repair module and a state inference module that uses system calls. Our evaluation shows that \name\ effectively repairs IaC scripts across multiple scenarios. By providing a flexible APR framework and a scalable evaluation methodology, our work establishes a foundation for future advancements in IaC repair. 
Future work will focus on seamlessly integrating it into sysadmins' workflows by enabling it to operate in the background, automatically suggesting fixes, for instance, through pull requests. We will also extend \name\ with more repair and state inference modules, and support for additional technologies and resource types.

\begin{acks}
We would like to thank Margarida Ferreira for her help with string encoding, Aaron Weiss and Arjun Guha for their assistance with Tortoise, Vasco Manquinho and Inês Lynce for their feedback, and Bruno Saavedra for his help with the video.
This work was supported by Fundação para a Ciência e a Tecnologia (FCT): N. Saavedra by grant BD/04736/2023;
N. Saavedra and J.\,F. Ferreira by
projects UIDB/50021/2020 (DOI: 10.54499/UIDB/50021/2020) and the `InfraGov' project, with ref. n. 2024.07411.IACDC (DOI: 10.54499/2024.07411.IACDC), funded by the `Plano de Recuperação e Resiliência (PRR)' under the investment `RE-C05-i08 - Ciência Mais Digital', measure `RE-C05-i08.M04' (in accordance with the FCT Notice No. 04/C05 i08/2024), framed within the financing agreement signed between the `Estrutura de Missão Recuperar Portugal (EMRP)' and the FCT as an intermediary beneficiary.
A. Mendes was financed by National Funds through the Portuguese funding agency, FCT, within project LA/P/0063/2020 (DOI: 10.54499/LA/P/0063/2020).

\end{acks}

\bibliographystyle{ACM-Reference-Format}
\bibliography{references}


\begin{thebibliography}{44}


\ifx \showCODEN    \undefined \def \showCODEN     #1{\unskip}     \fi
\ifx \showDOI      \undefined \def \showDOI       #1{#1}\fi
\ifx \showISBNx    \undefined \def \showISBNx     #1{\unskip}     \fi
\ifx \showISBNxiii \undefined \def \showISBNxiii  #1{\unskip}     \fi
\ifx \showISSN     \undefined \def \showISSN      #1{\unskip}     \fi
\ifx \showLCCN     \undefined \def \showLCCN      #1{\unskip}     \fi
\ifx \shownote     \undefined \def \shownote      #1{#1}          \fi
\ifx \showarticletitle \undefined \def \showarticletitle #1{#1}   \fi
\ifx \showURL      \undefined \def \showURL       {\relax}        \fi
\providecommand\bibfield[2]{#2}
\providecommand\bibinfo[2]{#2}
\providecommand\natexlab[1]{#1}
\providecommand\showeprint[2][]{arXiv:#2}

\bibitem[{Ansible project contributors}(2025)]%
        {AnsibleLighsail_2025}
\bibfield{author}{\bibinfo{person}{{Ansible project contributors}}.} \bibinfo{year}{2025}\natexlab{}.
\newblock \bibinfo{title}{{community.aws.lightsail module}}.
\newblock
\newblock
\urldef\tempurl%
\url{https://docs.ansible.com/ansible/latest/collections/community/aws/lightsail\_module.html}
\showURL{%
\tempurl}
\newblock
\shownote{[Accessed 20-03-2025]}.


\bibitem[Asher-Schapiro and Teixeira(2021)]%
        {facebook_down_2021}
\bibfield{author}{\bibinfo{person}{Avi Asher-Schapiro} {and} \bibinfo{person}{Fabio Teixeira}.} \bibinfo{year}{2021}\natexlab{}.
\newblock \bibinfo{title}{{Facebook down: What the outage meant for the developing world}}.
\newblock
\newblock
\urldef\tempurl%
\url{https://news.trust.org/item/20211005204816-qzjft/}
\showURL{%
\tempurl}
\newblock
\shownote{[Accessed 04-09-2024]}.


\bibitem[Begoug et~al\mbox{.}(2024)]%
        {begoug2024fine}
\bibfield{author}{\bibinfo{person}{Mahi Begoug}, \bibinfo{person}{Moataz Chouchen}, \bibinfo{person}{Ali Ouni}, \bibinfo{person}{Eman Abdullah~Alomar}, {and} \bibinfo{person}{Mohamed~Wiem Mkaouer}.} \bibinfo{year}{2024}\natexlab{}.
\newblock \showarticletitle{Fine-grained just-in-time defect prediction at the block level in Infrastructure-as-Code (IaC)}. In \bibinfo{booktitle}{\emph{Proceedings of the 21st International Conference on Mining Software Repositories}}. \bibinfo{pages}{100--112}.
\newblock


\bibitem[Bessghaier et~al\mbox{.}(2024)]%
        {bessghaier2024prevalence}
\bibfield{author}{\bibinfo{person}{Narjes Bessghaier}, \bibinfo{person}{Mahi Begoug}, \bibinfo{person}{Chemseddine Mebarki}, \bibinfo{person}{Ali Ouni}, \bibinfo{person}{Mohammed Sayagh}, {and} \bibinfo{person}{Mohamed~Wiem Mkaouer}.} \bibinfo{year}{2024}\natexlab{}.
\newblock \showarticletitle{On the prevalence, co-occurrence, and impact of infrastructure-as-code smells}. In \bibinfo{booktitle}{\emph{2024 IEEE International Conference on Software Analysis, Evolution and Reengineering (SANER)}}. IEEE, \bibinfo{pages}{23--34}.
\newblock


\bibitem[Bui et~al\mbox{.}(2023)]%
        {bui2023dockercleaner}
\bibfield{author}{\bibinfo{person}{Quang-Cuong Bui}, \bibinfo{person}{Malte Lauk{\"o}tter}, {and} \bibinfo{person}{Riccardo Scandariato}.} \bibinfo{year}{2023}\natexlab{}.
\newblock \showarticletitle{DockerCleaner: Automatic Repair of Security Smells in Dockerfiles}. In \bibinfo{booktitle}{\emph{2023 IEEE International Conference on Software Maintenance and Evolution}}. IEEE, \bibinfo{pages}{160--170}.
\newblock


\bibitem[Durieux(2024)]%
        {durieux2023parfum}
\bibfield{author}{\bibinfo{person}{Thomas Durieux}.} \bibinfo{year}{2024}\natexlab{}.
\newblock \showarticletitle{Empirical study of the docker smells impact on the image size}. In \bibinfo{booktitle}{\emph{Proceedings of the IEEE/ACM 46th Int. Conf. on Software Engineering}}. \bibinfo{pages}{1--12}.
\newblock


\bibitem[{GitLab}(2014)]%
        {gitlab_outage_2014}
\bibfield{author}{\bibinfo{person}{{GitLab}}.} \bibinfo{year}{2014}\natexlab{}.
\newblock \bibinfo{title}{{Gitlab.com Downtime 2014-07-07 Postmortem}}.
\newblock
\newblock
\urldef\tempurl%
\url{https://docs.google.com/document/d/1ScqXAdb6BjhsDzCo3qdPYbt1uULzgZqPO8zHeHHarS0}
\showURL{%
\tempurl}
\newblock
\shownote{[Accessed 04-09-2024]}.


\bibitem[Gonçalves(2023)]%
        {goncalves2023}
\bibfield{author}{\bibinfo{person}{João Gonçalves}.} \bibinfo{year}{2023}\natexlab{}.
\newblock \emph{\bibinfo{title}{{Automatic Detection of Security Smells in Terraform}}}.
\newblock Master's thesis. \bibinfo{school}{Instituto Superior Técnico}.
\newblock
\newblock
\shownote{Available at \url{https://fenix.tecnico.ulisboa.pt/cursos/meic-a/dissertacao/1972678479055768}}.


\bibitem[{gooeyblob}(2016)]%
        {reddit_outage_2016}
\bibfield{author}{\bibinfo{person}{{gooeyblob}}.} \bibinfo{year}{2016}\natexlab{}.
\newblock \bibinfo{title}{{Why Reddit was down on Aug 11}}.
\newblock
\newblock
\urldef\tempurl%
\url{https://web.archive.org/web/20221029203405/https://www.reddit.com/r/announcements/comments/4y0m56/why_reddit_was_down_on_aug_11/}
\showURL{%
\tempurl}
\newblock
\shownote{[Accessed 04-09-2024]}.


\bibitem[Goues et~al\mbox{.}(2019)]%
        {goues2019automated}
\bibfield{author}{\bibinfo{person}{Claire~Le Goues}, \bibinfo{person}{Michael Pradel}, {and} \bibinfo{person}{Abhik Roychoudhury}.} \bibinfo{year}{2019}\natexlab{}.
\newblock \showarticletitle{Automated program repair}.
\newblock \bibinfo{journal}{\emph{Commun. ACM}} \bibinfo{volume}{62}, \bibinfo{number}{12} (\bibinfo{year}{2019}), \bibinfo{pages}{56--65}.
\newblock


\bibitem[Guerriero et~al\mbox{.}(2019)]%
        {guerriero2019adoption}
\bibfield{author}{\bibinfo{person}{Michele Guerriero}, \bibinfo{person}{Martin Garriga}, \bibinfo{person}{Damian~A Tamburri}, {and} \bibinfo{person}{Fabio Palomba}.} \bibinfo{year}{2019}\natexlab{}.
\newblock \showarticletitle{Adoption, support, and challenges of infrastructure-as-code: Insights from industry}. In \bibinfo{booktitle}{\emph{2019 IEEE Int. Conf. on Software Maintenance and Evolution}}. \bibinfo{pages}{580--589}.
\newblock


\bibitem[Hasan et~al\mbox{.}(2020)]%
        {hasan2020testing}
\bibfield{author}{\bibinfo{person}{Mohammed~Mehedi Hasan}, \bibinfo{person}{Farzana~Ahamed Bhuiyan}, {and} \bibinfo{person}{Akond Rahman}.} \bibinfo{year}{2020}\natexlab{}.
\newblock \showarticletitle{Testing practices for infrastructure as code}. In \bibinfo{booktitle}{\emph{Proceedings of the 1st ACM SIGSOFT International Workshop on Languages and Tools for Next-Generation Testing}}. \bibinfo{pages}{7--12}.
\newblock


\bibitem[{HashiCorp}(2025)]%
        {TerraformLighsail_2025}
\bibfield{author}{\bibinfo{person}{{HashiCorp}}.} \bibinfo{year}{2025}\natexlab{}.
\newblock \bibinfo{title}{{Resource: aws\_lightsail\_instance}}.
\newblock
\newblock
\urldef\tempurl%
\url{https://registry.terraform.io/providers/hashicorp/aws/latest/docs/resources/lightsail\_instance}
\showURL{%
\tempurl}
\newblock
\shownote{[Accessed 20-03-2025]}.


\bibitem[Hassan et~al\mbox{.}(2018)]%
        {hassan2018rudsea}
\bibfield{author}{\bibinfo{person}{Foyzul Hassan}, \bibinfo{person}{Rodney Rodriguez}, {and} \bibinfo{person}{Xiaoyin Wang}.} \bibinfo{year}{2018}\natexlab{}.
\newblock \showarticletitle{Rudsea: recommending updates of dockerfiles via software environment analysis}. In \bibinfo{booktitle}{\emph{Proceedings of the 33rd ACM/IEEE International Conference on Automated Software Engineering}}.
\newblock


\bibitem[Hassan et~al\mbox{.}(2024)]%
        {hassan2024state}
\bibfield{author}{\bibinfo{person}{Md~Mahadi Hassan}, \bibinfo{person}{John Salvador}, \bibinfo{person}{Shubhra Kanti~Karmaker Santu}, {and} \bibinfo{person}{Akond Rahman}.} \bibinfo{year}{2024}\natexlab{}.
\newblock \showarticletitle{State Reconciliation Defects in Infrastructure as Code}.
\newblock \bibinfo{journal}{\emph{Proceedings of the ACM on Software Engineering}} \bibinfo{volume}{1}, \bibinfo{number}{FSE} (\bibinfo{year}{2024}).
\newblock


\bibitem[Henkel et~al\mbox{.}(2021)]%
        {henkel2021shipwright}
\bibfield{author}{\bibinfo{person}{Jordan Henkel}, \bibinfo{person}{Denini Silva}, \bibinfo{person}{Leopoldo Teixeira}, \bibinfo{person}{Marcelo d’Amorim}, {and} \bibinfo{person}{Thomas Reps}.} \bibinfo{year}{2021}\natexlab{}.
\newblock \showarticletitle{Shipwright: A human-in-the-loop system for dockerfile repair}. In \bibinfo{booktitle}{\emph{2021 IEEE/ACM 43rd International Conference on Software Engineering (ICSE)}}.
\newblock


\bibitem[{Honeycomb}(2021)]%
        {incident_resolution_honeycomb_2021}
\bibfield{author}{\bibinfo{person}{{Honeycomb}}.} \bibinfo{year}{2021}\natexlab{}.
\newblock \bibinfo{title}{{Incident Resolution: Do You Remember, the Twenty Fires of September?}}
\newblock
\newblock
\urldef\tempurl%
\url{https://www.honeycomb.io/blog/incident-resolution-september-retrospective}
\showURL{%
\tempurl}
\newblock
\shownote{[Accessed 04-09-2024]}.


\bibitem[{Jan David Nose}(2023)]%
        {rust_outage_2023}
\bibfield{author}{\bibinfo{person}{{Jan David Nose}}.} \bibinfo{year}{2023}\natexlab{}.
\newblock \bibinfo{title}{{DNS Outage on 2023-01-25}}.
\newblock
\newblock
\urldef\tempurl%
\url{https://blog.rust-lang.org/inside-rust/2023/02/08/dns-outage-portmortem.html}
\showURL{%
\tempurl}
\newblock
\shownote{[Accessed 04-09-2024]}.


\bibitem[Janardhan(2021)]%
        {Janardhan_2021}
\bibfield{author}{\bibinfo{person}{Santosh Janardhan}.} \bibinfo{year}{2021}\natexlab{}.
\newblock \bibinfo{title}{More details about the October 4 outage}.
\newblock
\newblock
\urldef\tempurl%
\url{https://engineering.fb.com/2021/10/05/networking-traffic/outage-details/}
\showURL{%
\tempurl}


\bibitem[{Khantanjil}(2025)]%
        {steam_issue_2025}
\bibfield{author}{\bibinfo{person}{{Khantanjil}}.} \bibinfo{year}{2025}\natexlab{}.
\newblock \bibinfo{title}{{Error: You are missing the following 32-bit libraries, and Steam may not run: libGL.so.1 libdrm.so.2}}.
\newblock
\newblock
\urldef\tempurl%
\url{https://github.com/ValveSoftware/steam-for-linux/issues/7284}
\showURL{%
\tempurl}
\newblock
\shownote{[Accessed 21-03-2025]}.


\bibitem[Lepiller et~al\mbox{.}(2021)]%
        {lepiller2021analyzing}
\bibfield{author}{\bibinfo{person}{Julien Lepiller}, \bibinfo{person}{Ruzica Piskac}, \bibinfo{person}{Martin Sch{\"a}f}, {and} \bibinfo{person}{Mark Santolucito}.} \bibinfo{year}{2021}\natexlab{}.
\newblock \showarticletitle{Analyzing infrastructure as code to prevent intra-update sniping vulnerabilities}. In \bibinfo{booktitle}{\emph{27th International Conference on Tools and Algorithms for the Construction and Analysis of Systems (TACAS)}}. Springer, \bibinfo{pages}{105--123}.
\newblock


\bibitem[Monperrus(2018a)]%
        {monperrus2018automatic}
\bibfield{author}{\bibinfo{person}{Martin Monperrus}.} \bibinfo{year}{2018}\natexlab{a}.
\newblock \showarticletitle{Automatic software repair: A bibliography}.
\newblock \bibinfo{journal}{\emph{ACM Computing Surveys (CSUR)}} \bibinfo{volume}{51}, \bibinfo{number}{1} (\bibinfo{year}{2018}), \bibinfo{pages}{1--24}.
\newblock


\bibitem[Monperrus(2018b)]%
        {repair-living-review}
\bibfield{author}{\bibinfo{person}{Martin Monperrus}.} \bibinfo{year}{2018}\natexlab{b}.
\newblock \bibinfo{booktitle}{\emph{The Living Review on Automated Program Repair}}.
\newblock \bibinfo{type}{{T}echnical {R}eport} hal-01956501. \bibinfo{institution}{HAL/archives-ouvertes.fr}.
\newblock


\bibitem[Opdebeeck et~al\mbox{.}(2022)]%
        {opdebeeck2022smelly}
\bibfield{author}{\bibinfo{person}{Ruben Opdebeeck}, \bibinfo{person}{Ahmed Zerouali}, {and} \bibinfo{person}{Coen De~Roover}.} \bibinfo{year}{2022}\natexlab{}.
\newblock \showarticletitle{Smelly variables in ansible infrastructure code: Detection, prevalence, and lifetime}. In \bibinfo{booktitle}{\emph{Proceedings of the 19th International Conference on Mining Software Repositories}}. \bibinfo{pages}{61--72}.
\newblock


\bibitem[Opdebeeck et~al\mbox{.}(2023)]%
        {opdebeeck2023control}
\bibfield{author}{\bibinfo{person}{Ruben Opdebeeck}, \bibinfo{person}{Ahmed Zerouali}, {and} \bibinfo{person}{Coen De~Roover}.} \bibinfo{year}{2023}\natexlab{}.
\newblock \showarticletitle{Control and data flow in security smell detection for infrastructure as code: Is it worth the effort?}. In \bibinfo{booktitle}{\emph{2023 IEEE/ACM 20th International Conference on Mining Software Repositories}}.
\newblock


\bibitem[{Puppet, Inc.}(2025)]%
        {array_titles_puppet_2025}
\bibfield{author}{\bibinfo{person}{{Puppet, Inc.}}} \bibinfo{year}{2025}\natexlab{}.
\newblock \bibinfo{title}{{Puppet Documentation - Resources: Arrays of Titles}}.
\newblock
\newblock
\urldef\tempurl%
\url{https://www.puppet.com/docs/puppet/latest/lang_resources.html\#lang\_resource\_syntax-arrays-of-titles}
\showURL{%
\tempurl}
\newblock
\shownote{[Accessed 21-03-2025]}.


\bibitem[Rahman et~al\mbox{.}(2019a)]%
        {rahman2019systematic}
\bibfield{author}{\bibinfo{person}{Akond Rahman}, \bibinfo{person}{Rezvan Mahdavi-Hezaveh}, {and} \bibinfo{person}{Laurie Williams}.} \bibinfo{year}{2019}\natexlab{a}.
\newblock \showarticletitle{A systematic mapping study of infrastructure as code research}.
\newblock \bibinfo{journal}{\emph{Information and Software Technology}}  \bibinfo{volume}{108} (\bibinfo{year}{2019}), \bibinfo{pages}{65--77}.
\newblock


\bibitem[Rahman et~al\mbox{.}(2019b)]%
        {rahman2019seven}
\bibfield{author}{\bibinfo{person}{Akond Rahman}, \bibinfo{person}{Chris Parnin}, {and} \bibinfo{person}{Laurie Williams}.} \bibinfo{year}{2019}\natexlab{b}.
\newblock \showarticletitle{The seven sins: Security smells in infrastructure as code scripts}. In \bibinfo{booktitle}{\emph{2019 IEEE/ACM 41st International Conference on Software Engineering (ICSE)}}. IEEE, \bibinfo{pages}{164--175}.
\newblock


\bibitem[Rahman et~al\mbox{.}(2021)]%
        {rahman2021security}
\bibfield{author}{\bibinfo{person}{Akond Rahman}, \bibinfo{person}{Md~Rayhanur Rahman}, \bibinfo{person}{Chris Parnin}, {and} \bibinfo{person}{Laurie Williams}.} \bibinfo{year}{2021}\natexlab{}.
\newblock \showarticletitle{Security smells in ansible and chef scripts: A replication study}.
\newblock \bibinfo{journal}{\emph{ACM Transactions on Software Engineering and Methodology (TOSEM)}} \bibinfo{volume}{30}, \bibinfo{number}{1} (\bibinfo{year}{2021}), \bibinfo{pages}{1--31}.
\newblock


\bibitem[Rahman et~al\mbox{.}(2023)]%
        {rahman2023security}
\bibfield{author}{\bibinfo{person}{Akond Rahman}, \bibinfo{person}{Shazibul~Islam Shamim}, \bibinfo{person}{Dibyendu~Brinto Bose}, {and} \bibinfo{person}{Rahul Pandita}.} \bibinfo{year}{2023}\natexlab{}.
\newblock \showarticletitle{Security misconfigurations in open source kubernetes manifests: An empirical study}.
\newblock \bibinfo{journal}{\emph{ACM Transactions on Software Engineering and Methodology}} \bibinfo{volume}{32}, \bibinfo{number}{4} (\bibinfo{year}{2023}), \bibinfo{pages}{1--36}.
\newblock


\bibitem[Reis et~al\mbox{.}(2022)]%
        {reis2022leveraging}
\bibfield{author}{\bibinfo{person}{Sofia Reis}, \bibinfo{person}{Rui Abreu}, \bibinfo{person}{Marcelo d'Amorim}, {and} \bibinfo{person}{Daniel Fortunato}.} \bibinfo{year}{2022}\natexlab{}.
\newblock \showarticletitle{Leveraging practitioners’ feedback to improve a security linter}. In \bibinfo{booktitle}{\emph{Proceedings of the 37th IEEE/ACM International Conference on Automated Software Engineering}}. \bibinfo{pages}{1--12}.
\newblock


\bibitem[Saavedra and Ferreira(2022)]%
        {saavedra2022glitch}
\bibfield{author}{\bibinfo{person}{Nuno Saavedra} {and} \bibinfo{person}{Jo{\~a}o~F. Ferreira}.} \bibinfo{year}{2022}\natexlab{}.
\newblock \showarticletitle{{GLITCH: Automated Polyglot Security Smell Detection in Infrastructure as Code}}. In \bibinfo{booktitle}{\emph{Proceedings of the 37th IEEE/ACM International Conference on Automated Software Engineering}}.
\newblock


\bibitem[Saavedra et~al\mbox{.}(2023)]%
        {saavedra2023polyglot}
\bibfield{author}{\bibinfo{person}{Nuno Saavedra}, \bibinfo{person}{Jo{\~a}o Gon{\c{c}}alves}, \bibinfo{person}{Miguel Henriques}, \bibinfo{person}{Jo{\~a}o~F. Ferreira}, {and} \bibinfo{person}{Alexandra Mendes}.} \bibinfo{year}{2023}\natexlab{}.
\newblock \showarticletitle{{Polyglot Code Smell Detection for Infrastructure as Code with GLITCH}}. In \bibinfo{booktitle}{\emph{Proceedings of the 38th IEEE/ACM International Conference on Automated Software Engineering}}.
\newblock


\bibitem[Schwarz et~al\mbox{.}(2018)]%
        {schwarz2018code}
\bibfield{author}{\bibinfo{person}{Julian Schwarz}, \bibinfo{person}{Andreas Steffens}, {and} \bibinfo{person}{Horst Lichter}.} \bibinfo{year}{2018}\natexlab{}.
\newblock \showarticletitle{Code smells in infrastructure as code}. In \bibinfo{booktitle}{\emph{2018 11th International Conference on the Quality of Information and Communications Technology (QUATIC)}}. IEEE, \bibinfo{pages}{220--228}.
\newblock


\bibitem[Sharma et~al\mbox{.}(2016)]%
        {sharma2016does}
\bibfield{author}{\bibinfo{person}{Tushar Sharma}, \bibinfo{person}{Marios Fragkoulis}, {and} \bibinfo{person}{Diomidis Spinellis}.} \bibinfo{year}{2016}\natexlab{}.
\newblock \showarticletitle{Does your configuration code smell?}. In \bibinfo{booktitle}{\emph{Proceedings of the 13th international conference on mining software repositories}}. \bibinfo{pages}{189--200}.
\newblock


\bibitem[Sokolowski et~al\mbox{.}(2024a)]%
        {sokolowski2024automated}
\bibfield{author}{\bibinfo{person}{Daniel Sokolowski}, \bibinfo{person}{David Spielmann}, {and} \bibinfo{person}{Guido Salvaneschi}.} \bibinfo{year}{2024}\natexlab{a}.
\newblock \showarticletitle{Automated Infrastructure as Code Program Testing}.
\newblock \bibinfo{journal}{\emph{IEEE TSE}} (\bibinfo{year}{2024}).
\newblock


\bibitem[Sokolowski et~al\mbox{.}(2024b)]%
        {sokolowski2024pipr}
\bibfield{author}{\bibinfo{person}{Daniel Sokolowski}, \bibinfo{person}{David Spielmann}, {and} \bibinfo{person}{Guido Salvaneschi}.} \bibinfo{year}{2024}\natexlab{b}.
\newblock \showarticletitle{The PIPr dataset of public infrastructure as code programs}. In \bibinfo{booktitle}{\emph{2024 IEEE/ACM 21st International Conference on Mining Software Repositories (MSR)}}. IEEE, \bibinfo{pages}{498--503}.
\newblock


\bibitem[Solar-Lezama et~al\mbox{.}(2006)]%
        {solar2006combinatorial}
\bibfield{author}{\bibinfo{person}{Armando Solar-Lezama}, \bibinfo{person}{Liviu Tancau}, \bibinfo{person}{Rastislav Bodik}, \bibinfo{person}{Sanjit Seshia}, {and} \bibinfo{person}{Vijay Saraswat}.} \bibinfo{year}{2006}\natexlab{}.
\newblock \showarticletitle{Combinatorial sketching for finite programs}. In \bibinfo{booktitle}{\emph{ASPLOS}}. \bibinfo{pages}{404--415}.
\newblock


\bibitem[Sotiropoulos et~al\mbox{.}(2020)]%
        {sotiropoulos2020practical}
\bibfield{author}{\bibinfo{person}{Thodoris Sotiropoulos}, \bibinfo{person}{Dimitris Mitropoulos}, {and} \bibinfo{person}{Diomidis Spinellis}.} \bibinfo{year}{2020}\natexlab{}.
\newblock \showarticletitle{Practical fault detection in puppet programs}. In \bibinfo{booktitle}{\emph{Proceedings of the ACM/IEEE 42nd International Conference on Software Engineering}}. \bibinfo{pages}{26--37}.
\newblock


\bibitem[{Stack Exchange Network}(2014)]%
        {stack_exchange_outage_2014}
\bibfield{author}{\bibinfo{person}{{Stack Exchange Network}}.} \bibinfo{year}{2014}\natexlab{}.
\newblock \bibinfo{title}{{Outage Post-Mortem: August 25th, 2014}}.
\newblock
\newblock
\urldef\tempurl%
\url{https://web.archive.org/web/20201020103424/https://stackstatus.net/post/96025967369/outage-post-mortem-august-25th-2014}
\showURL{%
\tempurl}
\newblock
\shownote{[Accessed 04-09-2024]}.


\bibitem[{Stack Overflow}(2024)]%
        {Overflow_2024}
\bibfield{author}{\bibinfo{person}{{Stack Overflow}}.} \bibinfo{year}{2024}\natexlab{}.
\newblock \bibinfo{title}{{Stack Overflow Developer Survey 2024}}.
\newblock
\newblock
\urldef\tempurl%
\url{https://survey.stackoverflow.co/2024/}
\showURL{%
\tempurl}
\newblock
\shownote{[Accessed 03-09-2024]}.


\bibitem[Weiss et~al\mbox{.}(2017)]%
        {weiss2017tortoise}
\bibfield{author}{\bibinfo{person}{Aaron Weiss}, \bibinfo{person}{Arjun Guha}, {and} \bibinfo{person}{Yuriy Brun}.} \bibinfo{year}{2017}\natexlab{}.
\newblock \showarticletitle{Tortoise: Interactive system configuration repair}. In \bibinfo{booktitle}{\emph{2017 32nd IEEE/ACM International Conference on Automated Software Engineering (ASE)}}. IEEE, \bibinfo{pages}{625--636}.
\newblock


\bibitem[{Wikimedia}(2017)]%
        {wikimedia_dataloss_2016}
\bibfield{author}{\bibinfo{person}{{Wikimedia}}.} \bibinfo{year}{2017}\natexlab{}.
\newblock \bibinfo{title}{{Incidents/2017-01-18 Labs}}.
\newblock
\newblock
\urldef\tempurl%
\url{https://wikitech.wikimedia.org/wiki/Incidents/2017-01-18_Labs}
\showURL{%
\tempurl}
\newblock
\shownote{[Accessed 04-09-2024]}.


\bibitem[Zerouali et~al\mbox{.}(2023)]%
        {zerouali2023helm}
\bibfield{author}{\bibinfo{person}{Ahmed Zerouali}, \bibinfo{person}{Ruben Opdebeeck}, {and} \bibinfo{person}{Coen De~Roover}.} \bibinfo{year}{2023}\natexlab{}.
\newblock \showarticletitle{Helm charts for Kubernetes applications: Evolution, outdatedness and security risks}. In \bibinfo{booktitle}{\emph{2023 IEEE/ACM 20th International Conference on Mining Software Repositories (MSR)}}.
\newblock


\end{thebibliography}

\end{document}